\documentclass[twocolumn,aps,pra,superscriptaddress]{revtex4-2}

\usepackage{graphicx,braket,float}
\usepackage{amsmath}
\usepackage{amsfonts}
\usepackage{amsbsy}
\usepackage{xcolor}
\usepackage{soul}
\usepackage{bm}
\usepackage[normalem]{ulem}
\usepackage[unicode]{hyperref}
\hypersetup{
	unicode=true, 
	plainpages=false,
	colorlinks=true,
	linkcolor=blue,
	citecolor=blue,
	filecolor=blue,
	urlcolor=blue
}
\usepackage{physics}
\usepackage{bbold}
\usepackage{epstopdf}

\DeclareMathOperator{\impart}{\mathrm{Im}}

\begin{document}


\title{Berezinskii-Kosterlitz-Thouless transitions in a ferromagnetic superfluid: effects of axial magnetization}



\author{Andrew P. C. Underwood}
\affiliation{Department of Physics, Centre for Quantum Science, and Dodd-Walls Centre for Photonic and Quantum Technologies, University of Otago, Dunedin, New Zealand}

\author{Andrew J. Groszek}
\affiliation{ARC Centre of Excellence for Engineered Quantum Systems, School of Mathematics and Physics, University of Queensland, Saint Lucia QLD 4072, Australia}
\affiliation{ARC Centre of Excellence in Future Low-Energy Electronics Technologies, School of Mathematics and Physics, University of Queensland, Saint Lucia QLD 4072, Australia}

\author{Xiaoquan Yu}
\affiliation{Graduate School of China Academy of Engineering Physics, Beijing 100193, China}
\affiliation{Department of Physics, Centre for Quantum Science, and Dodd-Walls Centre for Photonic and Quantum Technologies, University of Otago, Dunedin, New Zealand}

\author{P. B. Blakie}
\affiliation{Department of Physics, Centre for Quantum Science, and Dodd-Walls Centre for Photonic and Quantum Technologies, University of Otago, Dunedin, New Zealand}

\author{L. A. Williamson}
\affiliation{ARC Centre of Excellence for Engineered Quantum Systems, School of Mathematics and Physics, University of Queensland, Saint Lucia QLD 4072, Australia}


\date{\today}

\begin{abstract}
An easy-plane ferromagnetic spin-1 Bose gas undergoes two Berezinskii-Kosterlitz-Thouless (BKT) transitions, associated with mass and spin superfluidity respectively. We study the effect of axial magnetization on the superfluid properties of this system. We find that nonzero axial magnetization couples mass and spin superflow, via a mechanism analogous to the Andreev-Bashkin effect present in two-component superfluids. With sufficiently large axial magnetization mass and spin superfluidity arise simultaneously. The cross-over to this phase provides a finite-temperature generalization of the zero-temperature broken-axisymmetric to easy-axis transition. We present analytic relations connecting mass and spin superfluidity with experimentally observable coherence of the three spinor components and local magnetization.
\end{abstract}


\maketitle


\section{Introduction}

Spinor Bose gases possess a variety of ground-state phases that, in addition to global phase coherence, may exhibit nematic or ferromagnetic order \cite{PhysRevLett.81.742,doi:10.1143/JPSJ.67.1822,stenger_spin_1998,KAWAGUCHI2012253}. Consequently, such gases provide a rich platform to study equilibrium and non-equilibrium properties of both quantum and thermal phase transitions  \cite{sadler_spontaneous_2006,PhysRevA.84.063625,PhysRevLett.98.160404,PhysRevLett.99.130402,PhysRevA.76.043613,PhysRevLett.116.155301,barnett2011,PhysRevLett.116.025301,10.21468/SciPostPhys.7.3.029,PhysRevA.99.033611,PhysRevLett.131.183402,prufer_observation_2018,prufer_condensation_2022,PhysRevA.95.053638}. In a two-dimensional (2D) gas, thermal fluctuations preclude the formation of long-range order \cite{PhysRevLett.17.1133,PhysRev.158.383}, however systems may still exhibit Berezinskii-Kosterlitz-Thouless (BKT) type transitions~\cite{Berezinskii1970,Kosterlitz_1973,PhysRevLett.40.1727}. In spinor Bose gases, the interplay between spin and gauge symmetry can give rise to different BKT transitions and associated superfluid phases \cite{PhysRevLett.97.120406,PhysRevB.80.214513,PhysRevA.81.033616,PhysRevLett.106.140402,kobayashi2019,underwood2023}.

\begin{figure}[b]
	\includegraphics[width=\linewidth]{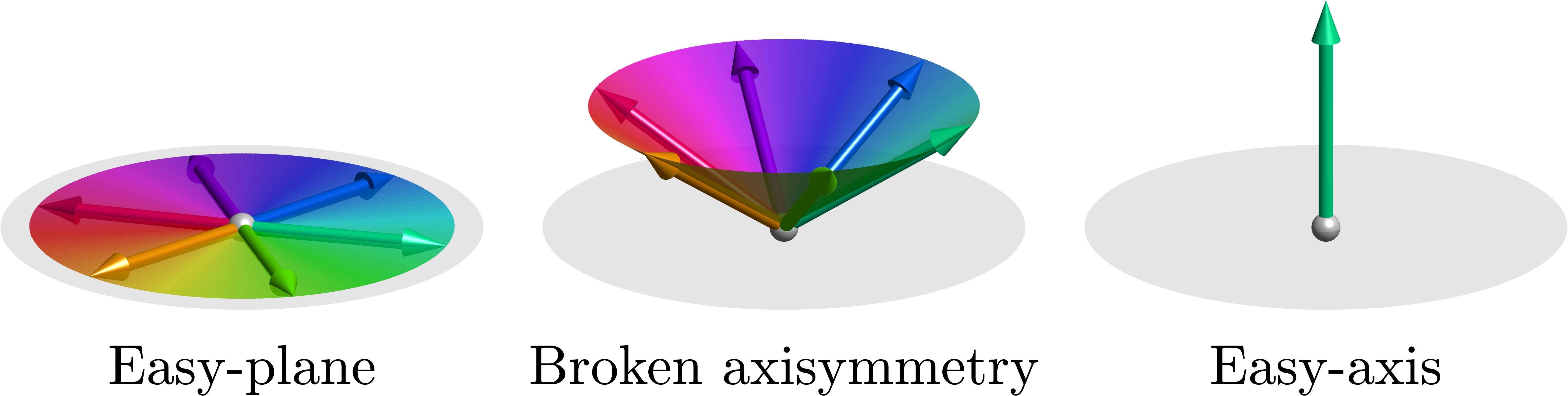}
	\caption{Schematic of zero-temperature ferromagnetic phases of a spin-1 Bose gas. At zero temperature, the transition from the broken axisymmetric to easy-axis phase occurs via a second order phase transition~\cite{murata2007}.\label{fig:schematic}}
\end{figure}

A spin-1 Bose gas in the easy-plane ferromagnetic phase exhibits distinct superflow of both mass and spin currents, corresponding to $\mathrm{U}(1)$ gauge and $\mathrm{SO}(2)$ spin-rotational symmetries respectively \cite{PhysRevA.95.053607,PhysRevB.97.224517}. The BKT transitions of this system have been studied previously; as temperature is decreased one observes first the emergence of mass superfluidity, followed at lower temperature by the emergence of spin superfluidity \cite{kobayashi2019,underwood2023}. As the (conserved) axial magnetization is increased the local magnetization tilts out of the plane, see Fig.~\ref{fig:schematic}. Ultimately, the gas becomes axially magnetized, losing spin-rotational symmetry. Here, in the easy-axis-ferromagnetic phase, superfluidity is akin to that of a single-component superfluid, with any superflow corresponding solely to the $\mathrm{U}(1)$ gauge symmetry. The nature of the superfluidity as the system transitions from an easy-plane through to an easy-axis-ferromagnetic system has not been explored.

In this paper we investigate the effect of increasing axial magnetization on the superfluid properties of a ferromagnetic 2D spin-1 Bose gas. We utilize a stochastic Gross-Pitaevskii model \cite{Gardiner_2002,Gardiner_2003,PhysRevA.77.033616,doi:10.1080/00018730802564254,PhysRevA.90.023631}, controlling the axial magnetization via the inclusion of a nonzero magnetic potential. With nonzero axial magnetization, mass and spin superfluidity are no longer independent quantities due to spin-gauge coupling~\cite{PhysRevLett.81.742,ho1996}. This motivates the introduction of a third superfluid quantity simultaneously describing superflow of mass and spin currents, analogous to the superfluid drag present in two-component superfluids \cite{andreev1976,PhysRevA.72.013616,Nespolo_2017,PhysRevResearch.3.023196}. We find that the spin superfluid transition temperature increases with increasing axial magnetization, eventually coinciding with the mass superfluid transition temperature. This equality of mass and spin superfluid transition temperatures occurs while the gas possesses only partial axial magnetization. We connect superfluidity with coherence in the individual spin components and in the transverse spin, finding analytic relations between superfluid densities and the algebraic decay of correlations. Our results give insight into the rich superfluid behavior possible in spinor systems and pave the way for experimental observation.

\section{Formalism}

\subsection{System}

We consider a 2D spin-1 Bose gas with low-energy macroscopically occupied modes described by a three-component classical field $\Psi=\left[\psi_{1},\psi_{0},\psi_{-1}\right]^{\mathrm{T}}$. The components $\psi_{m}$ denote the amplitudes of the $m\in\left\{1,0,-1\right\}$ magnetic sub-levels. The system energy is~\cite{PhysRevLett.81.742,doi:10.1143/JPSJ.67.1822}
\begin{equation}
	E = \int\dd{\mathbf{r}}\left[\Psi^{\dagger}\left(-\frac{\hbar^{2}\nabla^{2}}{2M}\mathbb{1}+qf_{z}^{2}\right)\Psi+\frac{g_{n}}{2}n^{2}+\frac{g_{s}}{2}|\mathbf{F}|^{2}\right]\label{eq:energy}
\end{equation}
where $M$ denotes the atomic mass and $q$ describes a uniform quadratic Zeeman shift arising from an external field along the $z$-spin axis~\cite{RevModPhys.85.1191}. The interactions are comprised of both density ($n=\Psi^{\dagger}\Psi$) and spin ($\mathbf{F}=\Psi^{\dagger}\mathbf{f}\Psi$) dependent terms, with respective 2D coupling constants $g_{n}>0$ and $g_{s}$. Here $\mathbf{f}=\left(f_{x},f_{y},f_{z}\right)$ denotes the vector of spin-1 matrices. We consider ferromagnetic interactions $g_{s}<0$, as realized in ultracold gases of $^{87}$Rb~\cite{Schmaljohann2004,Chang2004a} and $^7$Li~\cite{huh2020}. A linear Zeeman shift $p$ has been omitted from Eq.~(\ref{eq:energy}), as this can be removed via the transformation $\Psi\to\mathrm{e}^{-\mathrm{i}pf_{z}t/\hbar}\Psi$.


The total energy is invariant under both the global phase shift $\Psi\to\mathrm{e}^{\mathrm{i}\theta}\Psi$ and $z$-spin rotation $\Psi\to\mathrm{e}^{\mathrm{i}f_{z}\alpha}\Psi$. The corresponding conserved quantities are particle number $N=\int\dd{\mathbf{r}}\Psi^{\dagger}\Psi$ and $z$-magnetization $M_{z}=\int\dd{\mathbf{r}}\Psi^{\dagger}f_{z}\Psi$, with associated chemical and magnetic potentials $\mu$ and $\lambda$, respectively. The zero-temperature phases of the system in the mean-field regime are the ground states of Eq.~(\ref{eq:energy}). Here we focus on the ferromagnetic phases, which are shown in Fig.~\ref{fig:schematic}~\cite{murata2007}. With $\lambda=0$ and $0<q<2|g_{s}|n$ the axial magnetization is zero. Here, in addition to the breaking of $\mathrm{U}(1)$ gauge symmetry, the ground state breaks the $\mathrm{SO}(2)$ spin-rotational symmetry via the development of local magnetization transverse to the applied external field (``easy-plane''). The effect of $0<\lambda<q$ is the development of an axial component $F_{z}$ of the local magnetization, which depends on $\lambda$ and $q$ as~\cite{murata2007}
\begin{equation}
	\frac{F_{z}}{n}= \frac{\lambda\left(\lambda^{2}-q^{2}+2|g_{s}|nq\right)}{2|g_{s}|nq^{2}}.
\end{equation}
For $\lambda\geq q$ the system is in the easy-axis ferromagnetic phase. Here the ground state is axially magnetized $F_{z}=n$, and no longer breaks spin-rotational symmetry.

Although thermal fluctuations preclude symmetry breaking at nonzero temperature, quasi-long-range order below a BKT transition is still possible. In the easy-plane ferromagnetic gas ($\lambda=0$) quasi-long-range order can be present in both global phase and transverse spin, arising via distinct BKT transitions. Consequently, this system exhibits superfluid flow of both mass and spin currents. In this paper we detail the superfluid properties of the ferromagnetic spin-1 Bose gas as $\lambda$ is varied, determining the effects of nonzero axial magnetization.

\subsection{Theory of superfluidity with axial magnetization}

The superfluid properties of the spin-1 Bose gas may be evaluated by considering the system response to the combined global phase twist and $z$-spin rotation
\begin{equation}
\Psi\to \mathrm{e}^{\mathrm{i}\kappa_{n}\hat{\mathbf{n}}\cdot\mathbf{r}}\mathrm{e}^{\mathrm{i}f_{z}\kappa_{s}\hat{\mathbf{n}}\cdot\mathbf{r}}\Psi\label{eq:transform}.
\end{equation}
Here $\hat{\mathbf{n}}$ is a unit vector defining the twist direction. This transformation modifies the kinetic energy of the gas while leaving the remaining energy terms unchanged. The transformation~\eqref{eq:transform} acting on an equilibrium gas transforms the free energy to
\begin{equation}
	F = F_0+\int\dd{\mathbf{r}}\left(\rho_{nn}\frac{\hbar^{2}\kappa_{n}^{2}}{2M}+\rho_{ss}\frac{\hbar^{2}\kappa_{s}^{2}}{2M}+\rho_{ns}\frac{\hbar^{2}\kappa_{n}\kappa_{s}}{M}\right),\label{eq:dF}
\end{equation}
with $F_0$ the free energy prior to the transformation. In a system of dimensions $L\times L$ one has
\begin{equation}
	\rho_{ij}\equiv\frac{M}{\hbar^{2}L^{2}}\frac{\partial^{2}F}{\partial\kappa_{i}\partial\kappa_{j}}\bigg|_{\kappa_{i}=\kappa_{j}=0}\hspace{0.5cm}i,j\in\{n,s\}.\label{eq:sfdefns}
\end{equation}
The coefficients $\rho_{nn}$ and $\rho_{ss}$ define the mass and spin superfluid densities respectively.

The coefficient $\rho_{ns}=\rho_{sn}$ arises from interdependence of mass and spin currents in the presence of nonzero axial magnetization $\langle F_{z}\rangle $. To elucidate this, we write the free energy~\eqref{eq:dF} as a function of the superfluid velocities $\mathbf{v}_{i}(\mathbf{r})=\left(\hbar\kappa_{i}/M\right)\hat{\mathbf{n}}$. Taking functional derivatives of the free energy with respect to these velocities, one obtains the equilibrium superfluid currents as~\cite{svistunov_superfluid_2015}
\begin{equation}\label{eq:supercurrents}
\begin{split}
	\langle \mathbf{j}_n\rangle &=\frac{\delta F}{\delta \mathbf{v}_n}=\rho_{nn}\mathbf{v}_{n}+\rho_{ns}\mathbf{v}_{s},\\
	\langle \mathbf{j}_s\rangle &=\frac{\delta F}{\delta \mathbf{v}_s}=\rho_{ss}\mathbf{v}_{s}+\rho_{ns}\mathbf{v}_{n}.
\end{split}
\end{equation}
Motivated by Eq.~\eqref{eq:supercurrents} we interpret $\rho_{ns}$ as the portion of mass (spin) superfluid density that simultaneously contributes to spin (mass) superflow. Note $F-F_0\ge 0$ irrespective of $\kappa_n$ and $\kappa_s$, hence $\rho_{ns}\leq\sqrt{\rho_{nn}\rho_{ss}}$.  The coupling arises due to the inter-dependence of the two symmetry transformations, whereby spin rotations affect mass current and global phase rotations affect spin current. A quantity analogous to $\rho_{ns}$ arises in two-component superfluids, where it is termed the superfluid drag; this describes entrainment between the two components due to current-current coupling, known as the Andreev-Bashkin effect \cite{andreev1976}. Note $\rho_{ns}$ is distinct from the effect termed ``spin drag'' in Ref.~\cite{underwood2023}, which describes the tendency for component currents to entrain due to the spin exchange energy.
 
The total instantaneous mass and spin currents in the spinor system are, respectively~\cite{yukawa2012},
\begin{equation}\label{eq:totalcurrents}
\begin{split}
	\mathbf{J}_{n}&=\frac{\hbar}{M}\impart\left(\Psi^{\dagger}\nabla\Psi\right),\\
	\mathbf{J}_{s}&=\frac{\hbar}{M}\impart\left(\Psi^{\dagger}f_{z}\nabla\Psi\right).
\end{split}
\end{equation}
At zero temperature the total and superfluid currents are identical. Applying the transformation~\eqref{eq:transform} to the spatially uniform ground state and evaluating Eq.~\eqref{eq:totalcurrents} we identify:
\begin{equation}\label{eq:rhozeroT}
\begin{split}
\rho_{nn} &= n,\\
\rho_{ss} &= |\psi_1|^2+|\psi_{-1}|^2,\quad\text{(zero temperature)}\\
\rho_{ns} &= F_z,
\end{split}
\end{equation}
At zero temperature, the superfluid densities $\rho_{nn}$ and $\rho_{ns}$ are distinct in the broken-axis symmetric phase $\lambda<q$, but become equal in the easy-axis phase $\lambda\geq q$, where $|\psi_0|^2=|\psi_{-1}|^2=0$ and $F_z=n$.

\subsection{Model and simulation details}

We consider  a 2D spin-1 Bose gas confined to a box of dimensions $L\times L$, coupled to a grand-canonical reservoir with chemical potential $\mu$, magnetic potential $\lambda$, and temperature $T$. For a given quantity $\mathcal{O}$ equilibrium expectation values are given by
\begin{equation}
	\langle\mathcal{O}\rangle = \frac{1}{Z}\int\mathrm{D}\Psi\,\mathcal{O}\left[\Psi\right]\mathrm{e}^{-\left(E\left[\Psi\right]-\mu N\left[\Psi\right]-\lambda M_{z}\left[\Psi\right]\right)/k_{B}T},\label{eq:thermalexpectation}
\end{equation}
where $Z=\int\mathrm{D}\Psi\,\exp\left[-\left(E-\mu N-\lambda M_{z}\right)/k_{B}T\right]$ is the system partition function. In practice, we evaluate Eq.~(\ref{eq:thermalexpectation}) from stationary solutions of the stochastic spin-1 Gross-Pitaevskii equation, which sample the grand-canonical ensemble~\cite{doi:10.1080/00018730802564254} (also see Refs.~\cite{PhysRevA.100.033618,PhysRevResearch.4.033130}). This equation is:~\cite{Gardiner_2002,Gardiner_2003,PhysRevA.77.033616,PhysRevA.90.023631,doi:10.1080/00018730802564254}
\begin{equation}\label{eq:psidt}
	\mathrm{i}\hbar\dd{\Psi} = \left(1-\mathrm{i}\gamma\right)\left[\mathcal{L}\{\Psi\}-\left(\mu+\lambda f_{z}\right)\Psi\right]\dd{t}+\mathrm{i}\hbar\dd{W}.
\end{equation}
The nonlinear operator
\begin{equation}
	\mathcal{L}\{\Psi\} = \left(-\frac{\hbar^{2}\nabla^{2}}{2M}+qf_{z}+g_{n}n+g_{s}\sum_{\nu}F_{\nu}f_{\nu}\right)\Psi\label{eq:sgpe}
\end{equation}
describes time evolution due to kinetic energy, quadratic Zeeman shift, density-dependent interactions, and spin-dependent interactions ($\nu\in\left\{x,y,z\right\}$). The dimensionless parameter $\gamma$ describes the coupling strength between $\Psi$ and the grand canonical reservoir; stationary solutions are independent of $\gamma$. The components of $\dd{W}=\left[\mathrm{d}w_{1}, \mathrm{d}w_{0}, \mathrm{d}w_{-1}\right]^{T}$ are Gaussian-distributed complex noise with correlations $\expval{\mathrm{d}w_{m}(\mathbf{r})\mathrm{d}w_{m'}(\mathbf{r}')} = \left(2\gamma k_{B}T/\hbar\right)\tilde{\delta}(\mathbf{r},\mathbf{r}')\delta_{m,m'}\dd{t}$, where $\tilde{\delta}$ is a delta function in the space of macroscopically occupied modes \cite{doi:10.1080/00018730802564254}.

We perform simulations with periodic boundary conditions. We expand the field $\Psi$ in a plane-wave basis, with the constituent modes determined by our $\mathcal{N}\times \mathcal{N}$ point numerical grid. At large momenta interactions are unimportant, so that a mode of wavenumber $\mathbf{k}$ will have occupation $N_{\mathbf{k}}\approx 2Mk_{B}T/\hbar^{2}|\mathbf{k}|^{2}$. Motivated by this, we use grid spacing $\Delta x = \sqrt{2\pi\hbar^{2}/Mk_{B}T}$, enforcing occupation of at least order unity for all modes. We take $\mu$ as an energy scale with associated length $x_{\mu}=\hbar/\sqrt{M\mu}$, and compute dependence of equilibrium properties on the scaled temperature $\mathcal{T}=Mg_{n}k_{b}T/\hbar^{2}\mu$ and magnetic potential $\lambda/\mu$. Hereon, the remaining parameters are fixed as $Mg_{n}/\hbar^{2}=0.15$, $g_{s}=-0.1g_{n}$, $q=0.1\mu$, and $\gamma=0.1$, unless otherwise stated. We obtain stationary solutions of Eq.~(\ref{eq:psidt}) by evolving an initial state $\Psi=0$ for time $t\sim 10^{5}\hbar/\mu$, observing saturation of the zero momentum mode population to verify equilibration is reached. Following this, we obtain $\mathcal{N}_{\mathrm{s}}\sim 5\times 10^{4}$ samples $\Psi_i$ at intervals of $10\hbar/\mu$ and compute thermal expectation values via
\begin{equation}
	\langle\mathcal{O}\rangle\approx\frac{1}{\mathcal{N}_{\mathrm{s}}}\sum_{i=1}^{\mathcal{N}_{\mathrm{s}}} \mathcal{O}\left[\Psi_{i}\right].\label{eq:timeaverage}
\end{equation}


\section{Numerical results}

\subsection{Superfluid phases in the presence of axial magnetization}

\begin{figure}
	\includegraphics[width=\linewidth]{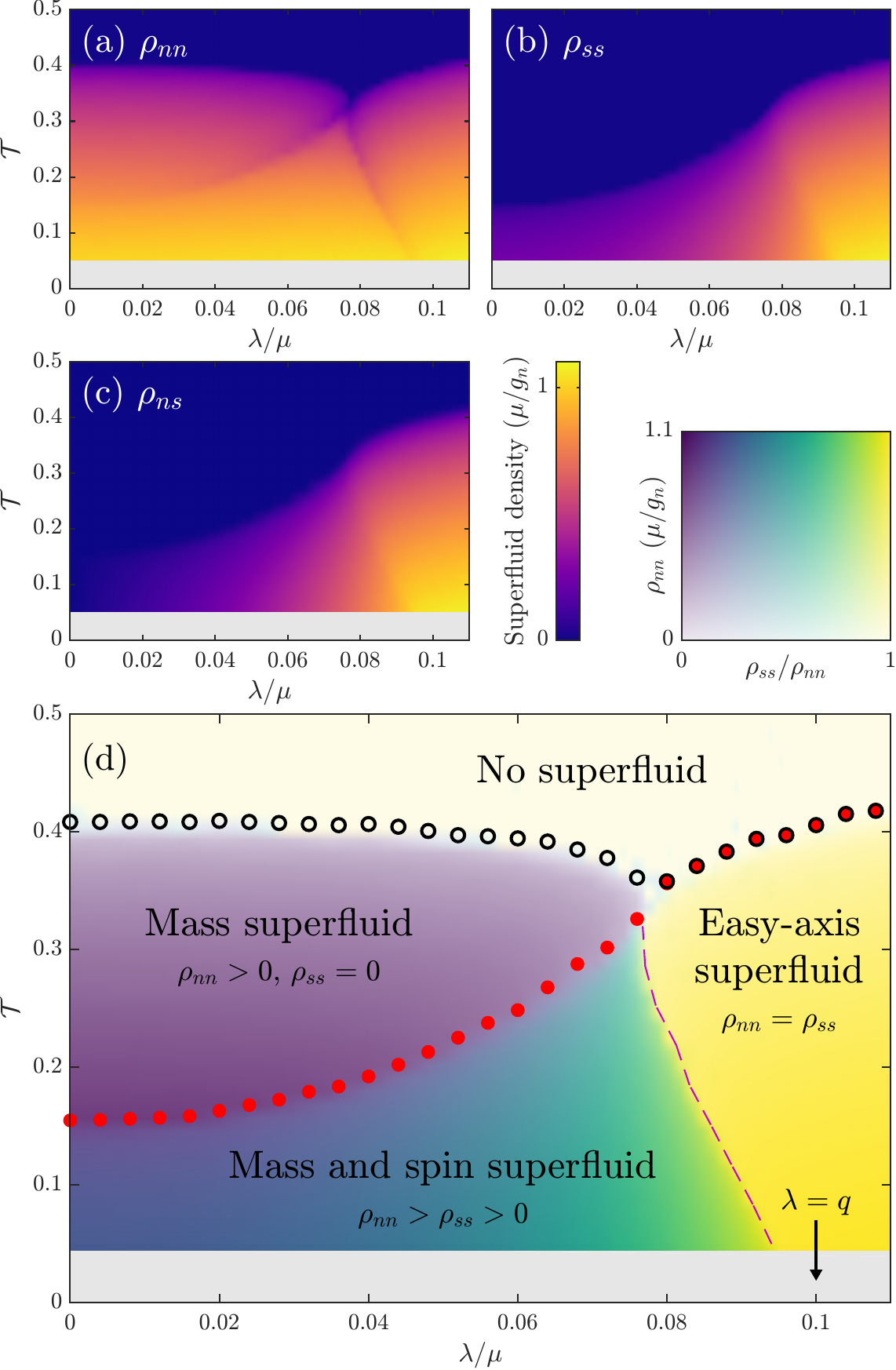}
	\caption{Superfluid densities (a) $\rho_{nn}$, (b) $\rho_{ss}$ and (c) $\rho_{ns}$ for varying temperature $\mathcal{T}$ and magnetic potential $\lambda$. (d) We identify three superfluid phases: first, a mass superfluid phase, with $\rho_{nn}>0$ and $\rho_{ss}=0$. Second, a mass and spin superfluid phase, with $\rho_{nn}>\rho_{ss}>0$. Third, an easy-axis superfluid phase, with $\rho_{nn}=\rho_{ss}=\rho_{ns}$. Black and red circles denote approximate mass and spin superfluid transition temperatures respectively, identified via the conditions $\rho_{nn}$, $\rho_{ss}>0.05\mu/g_{n}$. Purple dashed line denotes the curve $\lambda_{c}(\mathcal{T})$, identified via the condition $\rho_{nn}-\rho_{ss}<0.05\mu/g_{n}$. All results are computed with $\mathcal{N}=256$.\label{fig:superfluiddensities}}
\end{figure}



We first explore the change in superfluid properties of the system as the scaled temperature $\mathcal{T}$ and magnetic potential $\lambda$ are varied. Recent works~\cite{underwood2023,kobayashi2019} focusing on the unmagnetized case, $\lambda=0$, found that this system exhibits two distinct BKT transitions as it is cooled. It first transitions at temperature $\mathcal{T}_{n}$ to a phase exhibiting only mass superfluidity, before transitioning again at a lower temperature $\mathcal{T}_{s}$ to a phase where both spin and mass superfluidity coexist. Here we generalize these results to $\lambda > 0$.

The superfluid densities $\rho_{nn}$, $\rho_{ss}$, and $\rho_{ns}$, defined in Eq.~\eqref{eq:sfdefns}, can be extracted from equilibrium current-current correlations (see Appendix~\ref{computesuperfluid} for details). The dependence of these superfluid densities on both $\mathcal{T}$ and $\lambda$ is shown in Fig.~\ref{fig:superfluiddensities}(a)-(c). We observe three distinct superfluid phases. At low magnetization, the system behaves similarly to the $\lambda=0$ case: there is first a transition to a mass superfluid ($\rho_{nn} > 0$) at critical temperature $\mathcal{T}_n$ [panel (a)], followed by a second transition at a lower critical temperature $\mathcal{T}_s$, where spin superfluidity ($\rho_{ss} > 0$) also emerges [panel (b)]. As $\lambda$ is increased, the coupling $\rho_{ns}$ between the two superfluids grows [panel (c)]. The spin superfluid density $\rho_{ss}$ also grows with $\lambda$ until it coincides with $\rho_{nn}$ for $\lambda>\lambda_{c}(\mathcal{T})$, defining a third superfluid region. Here $\lambda_c(\mathcal{T})$ is the curve demarcating $\rho_{nn}>\rho_{ss}$ [$\lambda<\lambda_c(\mathcal{T})$] from $\rho_{nn}=\rho_{ss}$ [$\lambda\ge \lambda_c(\mathcal{T})$]. This cross-over generalizes the zero-temperature transition from the broken axisymmetry to the easy-axis phase. We therefore denote the region where $\rho_{nn}=\rho_{ss}$ an ``easy-axis superfluid''. The superfluid phase diagram is summarized in Fig.~\ref{fig:superfluiddensities}(d).

\begin{figure}
		\includegraphics[width=\linewidth]{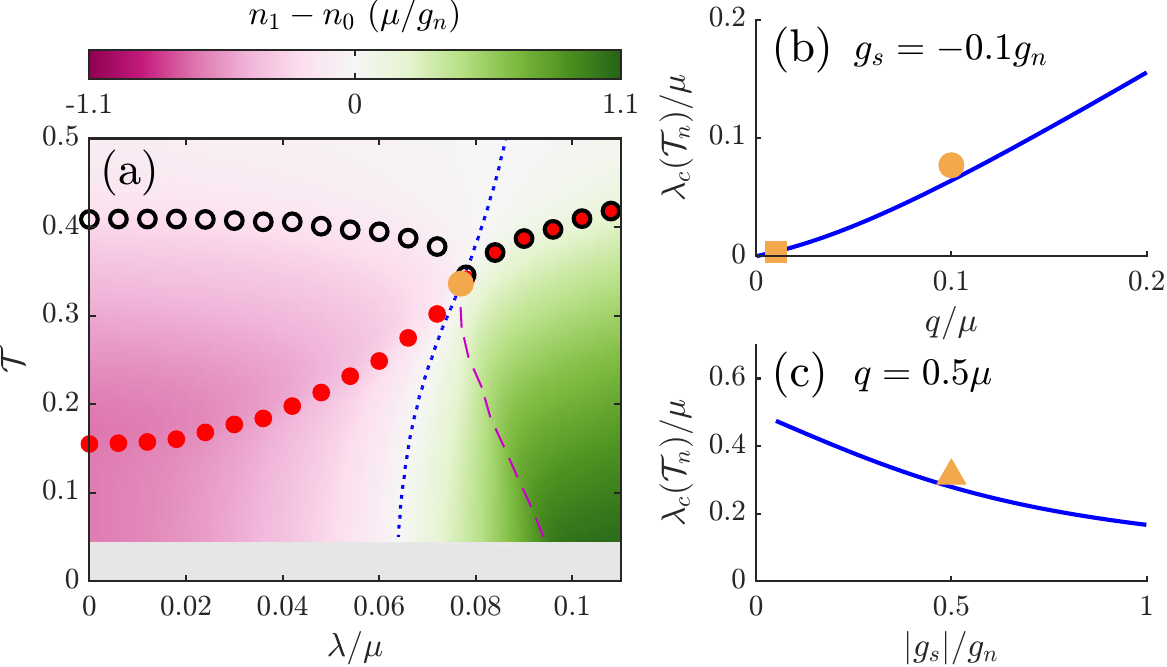}
		\caption{Role of component densities in determining $\lambda_{c}(\mathcal{T}_{n})$. (a) Density difference $n_1-n_0$. Black and red circles, and purple dashed line are as in Fig.~\ref{fig:superfluiddensities}(d). The mass and spin superfluid transitions coincide at $\lambda=\lambda_{c}(\mathcal{T}_{n})$ (orange circle). Dotted blue line is the curve $n_{1}= n_{0}$, which coincides with $\lambda_{c}(\mathcal{T}_{n})$ at $\mathcal{T}=\mathcal{T}_{n}$. (b) Dependence of $\lambda_{c}(\mathcal{T}_n)$ on quadratic Zeeman energy for $g_{s}=-0.1g_{n}$. (c) Dependence of $\lambda_{c}(\mathcal{T}_n)$ on $|g_{s}|$ with $q=0.5\mu$. In (b) and (c) blue lines are analytic estimates obtained from setting $n_1=n_0$ in the ground state, while orange markers are numerical results with $\left(g_{s}/g_{n},q/\mu\right) = \left(-0.1,0.1\right)$ (circle), $\left(-0.1,0.01\right)$ (square), and $\left(-0.5,0.5\right)$ (triangle). All results are computed with $\mathcal{N}=256$.\label{fig:compdens}}
	\end{figure}

The zero temperature value of $\lambda_{c}$ corresponds to the transition from the broken axisymmetry to the easy-axis ground state~\cite{murata2007}, i.e. $\lambda_{c}(0) = q$, see Fig.~\ref{fig:superfluiddensities}(d). At finite temperature, we find that $\lambda_{c}(\mathcal{T})<\lambda_{c}(0)$ and the gas has only partial axial magnetization. In particular, at $\mathcal{T}=\mathcal{T}_{n}$ we find $\lambda_{c}$ is determined by the condition that the component densities $n_{1} =\langle |\psi_{1}|^{2}\rangle$ and $n_{0}=\langle|\psi_{0}|^{2}\rangle$ are equal. In Fig.~\ref{fig:compdens}(a) we plot the difference $n_{1}-n_{0}$; the intersection of the curve $n_{1}=n_{0}$ (blue dotted line) with $\mathcal{T}_{n}$ (black circles) occurs at $\lambda\approx\lambda_{c}(\mathcal{T}_{n})$. Approximating $n_{0}$ and $n_{1}$ by their broken-axisymmetric ground state values gives an accurate analytic estimate for $\lambda_{c}(\mathcal{T}_{n})$ \footnote{For given $g_{n}$, $g_{s}$, $\mu$, and $q$ the estimate $\lambda_{c}(\mathcal{T}_{n})$ is determined as the positive root of the third-order polynomial in $\lambda$\begin{equation}
		\frac{q-\lambda}{q+\lambda} = \frac{1}{2}\frac{\lambda^{2}-q^{2}+2q|g_{s}|n(\lambda)}{\lambda^{2}+q^{2}+2q|g_{s}|n(\lambda)},
	\end{equation} where $n(\lambda)=\left(\mu-\frac{q}{2}+\frac{\lambda^{2}}{2q}\right)/\left(g_{n}-|g_{s}|\right)$.}, see Fig.~\ref{fig:compdens}. We have performed additional calculations of the phase diagram for $\left(g_{s}/g_{n},q/\mu\right)=\left(-0.1,0.01\right)$ and $\left(g_{s}/g_{n},q/\mu\right)=\left(-0.5,0.5\right)$. In both cases results are qualitatively similar to Fig.~\ref{fig:superfluiddensities}, with $\lambda_{c}(\mathcal{T}_{n})$ in agreement with the analytic estimate, see Fig.~\ref{fig:compdens}(b) and (c). 

\subsection{Phase coherence and correlations}\label{sec:coherence}

\begin{figure}
	\includegraphics[width=\linewidth]{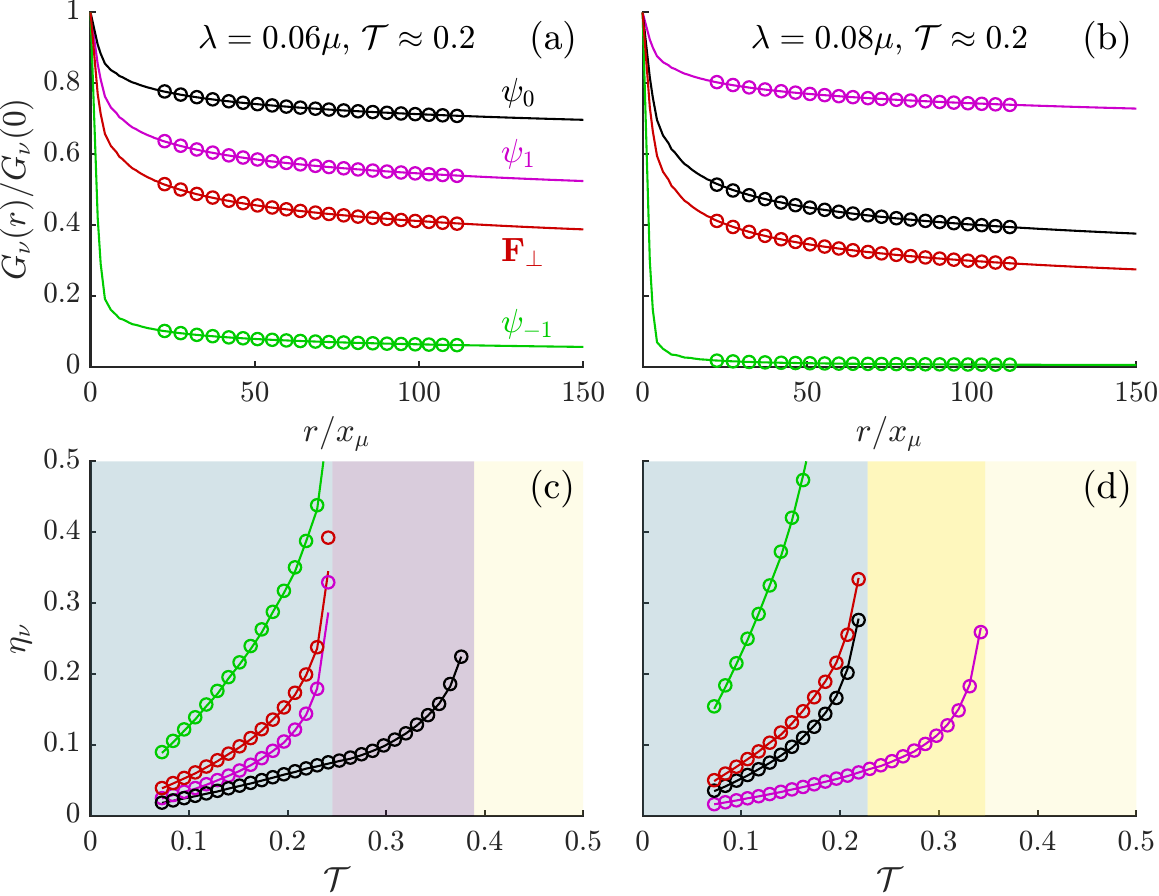}
	\caption{\label{Gfig} Top row: algebraically decaying correlations $G_{1}$ (purple), $G_{0}$ (black), $G_{-1}$ (green) and $G_{\perp}$ (red), at $\mathcal{T}\approx 0.2$ with (a) $\lambda=0.06\mu$ and (b) $\lambda=0.08\mu$. Lines are obtained from evaluation of Eq.~\eqref{eq:corr}. Circles are obtained from Eq.~\eqref{corrDecay} with decay exponents computed via superfluid densities (\ref{eq:eta_from_sf}), (\ref{eq:eaeta}) and (\ref{eq:meta}). Bottom row: comparison of decay exponents $\eta_\nu$ extracted from fitting to correlations (\ref{eq:corr}) (lines), and decay exponents computed from superfluid densities (\ref{eq:eta_from_sf}), (\ref{eq:eaeta}) and (\ref{eq:meta}) (circles), with (c) $\lambda=0.06\mu$ and (d) $\lambda=0.08\mu$. Colors are as in (a) and (b). Background colors indicate superfluid phases from Fig.~\ref{fig:superfluiddensities}(d). All results are computed with $\mathcal{N}=512$.}
\end{figure}

Superfluidity in two dimensions is associated with algebraic decay of correlations, signifying quasi-long-range order. In a spinor system, such order may be present in component and spin correlations
\begin{equation}\label{eq:corr}
\begin{split}
G_m(\mathbf{r})&=\langle \psi_m^*(\mathbf{r})\psi_m(\mathbf{0})\rangle,\\
G_\perp(\mathbf{r})&= \langle \mathbf{F}_\perp(\mathbf{r})\cdot\mathbf{F}_\perp(\mathbf{0})\rangle.
\end{split}
\end{equation}
At low temperature the behavior of these correlations is dominated by long-wavelength gapless modes~\cite{barnett2011}. This system exhibits two such modes, corresponding to the two symmetries in Eq.~\eqref{eq:transform}~\cite{uchino2010,symes2014}. Excitations of these modes are described via the energy functional
\begin{equation}\label{Hquadratic}
E=\frac{\hbar^2}{2M}\int\dd{\mathbf{r}} \left(\rho_{nn}|\nabla\theta|^2+\rho_{ss}|\nabla\alpha|^2+2\rho_{ns}\nabla\theta\cdot\nabla\alpha\right),
\end{equation}
where $\theta(\mathbf{r})$ and $\alpha(\mathbf{r})$ are the spatially varying global phase and transverse spin angle respectively. The effect of these excitations on the large $|\mathbf{r}|$ behavior of the correlations~\eqref{eq:corr} can be evaluated analytically from Eq.~\eqref{eq:thermalexpectation} using standard techniques~\cite{chaikin1995,wegner1967}. Within the mass and spin superfluid phase the result is
\begin{equation}\label{corrDecay}
	G_\nu(\mathbf{r})\propto |\mathbf{r}|^{-\eta_\nu},\quad\nu\in\left\{1,0,-1,\perp\right\},
\end{equation}
with
\begin{equation}
	\begin{split}
		\eta_{0} &= \frac{Mk_{B}T}{2\pi\hbar^{2}\rho_{nn}}\frac{1}{1-\mathcal{R}^2},\\
		\eta_{\perp} &= \frac{Mk_{B}T}{2\pi\hbar^{2}\rho_{ss}}\frac{1}{1-\mathcal{R}^2},\\
		\eta_{\pm 1} &= \eta_0+\eta_\perp\mp \frac{Mk_{B}T}{\pi\hbar^{2}\sqrt{\rho_{nn}\rho_{ss}}}\frac{\mathcal{R}}{1-\mathcal{R}^2}.
	\end{split}\label{eq:eta_from_sf}
\end{equation}
with $\mathcal{R}=\rho_{ns}/\sqrt{\rho_{nn}\rho_{ss}}$. Note that $\eta_{1}+\eta_{-1}=2\eta_{\perp}+2\eta_{0}$ and hence only three of the four exponents in Eq.~(\ref{eq:eta_from_sf}) are independent. Eq.~(\ref{eq:eta_from_sf}) can be written concisely as
\begin{equation}
	\eta = \frac{Mk_{B}T}{2\pi\hbar^{2}}\rho^{-1},\label{eq:conciseeta}
\end{equation}
where we have defined
\begin{equation}
	 \eta=\begin{bmatrix}
			\eta_{0} & -\bar{\eta}\\
			-\bar{\eta} & \eta_{\perp}
	\end{bmatrix},\quad \rho =  \begin{bmatrix}
		\rho_{nn} & \rho_{ns} \\
		\rho_{ns} & \rho_{ss}
	\end{bmatrix},
\end{equation}
with $\bar{\eta}=\left(\eta_{-1}-\eta_{1}\right)/4$.


In the easy-axis superfluid phase $\mathcal{R}=1$, and the expressions (\ref{eq:eta_from_sf}) are invalid. Rather, we note that with $\rho_{nn}=\rho_{ss}=\rho_{ns}$ Eq.~(\ref{Hquadratic}) reduces to $E=\left(\hbar^{2}/2M\right)\int\dd{\mathbf{r}}\rho_{ns}|\nabla\left(\theta+\alpha\right)|^{2}$, i.e. the energy reduces to that of a single-component field with fluctuations in the phase $\theta+\alpha$. Correspondingly, the only relevant correlation function is $G_{1}(\mathbf{r})\propto |\mathbf{r}|^{-\eta_{1}}$ with decay exponent satisfying
\begin{equation}
	\eta_{1} = \frac{Mk_{B}T}{2\pi\hbar^{2}\rho_{ns}}\quad\text{(Easy-axis superfluid).}\label{eq:eaeta}
\end{equation}
Similarly, in the mass superfluid phase $\rho_{ss}=\rho_{ns}=0$ and Eq.~(\ref{Hquadratic}) reduces to $E=\left(\hbar^{2}/2M\right)\int\dd{\mathbf{r}}\rho_{nn}|\nabla\theta|^{2}$. The relevant correlation function is then $G_{0}(\mathbf{r})\propto |\mathbf{r}|^{-\eta_{0}}$ with decay exponent
\begin{equation}
	\eta_{0} = \frac{Mk_{B}T}{2\pi\hbar^{2}\rho_{nn}}\quad\text{(Mass superfluid).}\label{eq:meta}
\end{equation}
These results connect quasi-long-range order with superfluidity. 

We have numerically determined the correlation functions (\ref{eq:corr}) using Eq.~(\ref{eq:timeaverage}). In the distinct mass and spin superfluid phase all $G_{\nu}$ exhibit algebraic decay. In the easy-axis superfluid phase only $G_{1}$ exhibits algebraic decay. In the mass superfluid phase only $G_{0}$ exhibits algebraic decay. Example correlations are shown in Fig.~\ref{Gfig}(a) and (b). We have also computed the long-range behavior of $G_{\nu}$ from the superfluid densities, using Eq.~(\ref{corrDecay}) with decay exponents as in Eqs.~(\ref{eq:eta_from_sf}), (\ref{eq:eaeta}) and (\ref{eq:meta}). This agrees with the correlations determined directly within the region $10\lambda_{\mathrm{th}}<|\mathbf{r}|<0.2L$, see Fig.~\ref{Gfig}(a) and (b). Here $\lambda_{\mathrm{th}}=\sqrt{2\pi\hbar^{2}/Mk_{B}T}$ is the thermal de Broglie wavelength; the upper bound of $0.2L$ is imposed as finite-size effects become significant as $|\mathbf{r}|\to L$.

Similarly, we have extracted the decay exponents $\eta_{\nu}$ from correlations computed via Eq.~(\ref{eq:timeaverage}) \footnote{We determine the presence of algebraic decay in each of the correlation functions (\ref{eq:corr}) by computing a quadratic fit $\ln\left[G_{\nu}^{\text{fit}}(r)\right]=a_{\nu}\ln^{2}(r)+b_{\nu}\ln(r)+c_{\nu}$ over the range $10\Delta x<r<0.2L$ at each temperature. If at some point $|a_{\nu}|<0.05$ we consider the correlations to exhibit algebraic decay. To extract the associated decay exponent $\eta_{\nu}$, we compute a linear fit $\ln\left[G_{\nu}^{\text{fit}}(r)\right]= -\eta_{\nu}\ln(r)+c_{\nu}$ over the same range.}, and compared these with the predictions (\ref{eq:eta_from_sf}), (\ref{eq:eaeta}) and (\ref{eq:meta}); see Fig.~\ref{Gfig}(c) and (d). We find agreement for all applicable temperatures. Interestingly, in the distinct mass and spin superfluid phase, where all $G_{\nu}$ exhibit algebraic decay, we find the decay exponents $\eta_{\nu}$ may exceed $1/4$. This is in contrast to the maximal decay exponent of a single-component 2D superfluid \cite{PhysRevLett.39.1201}.

\begin{figure}[t]
	\includegraphics[width=\linewidth]{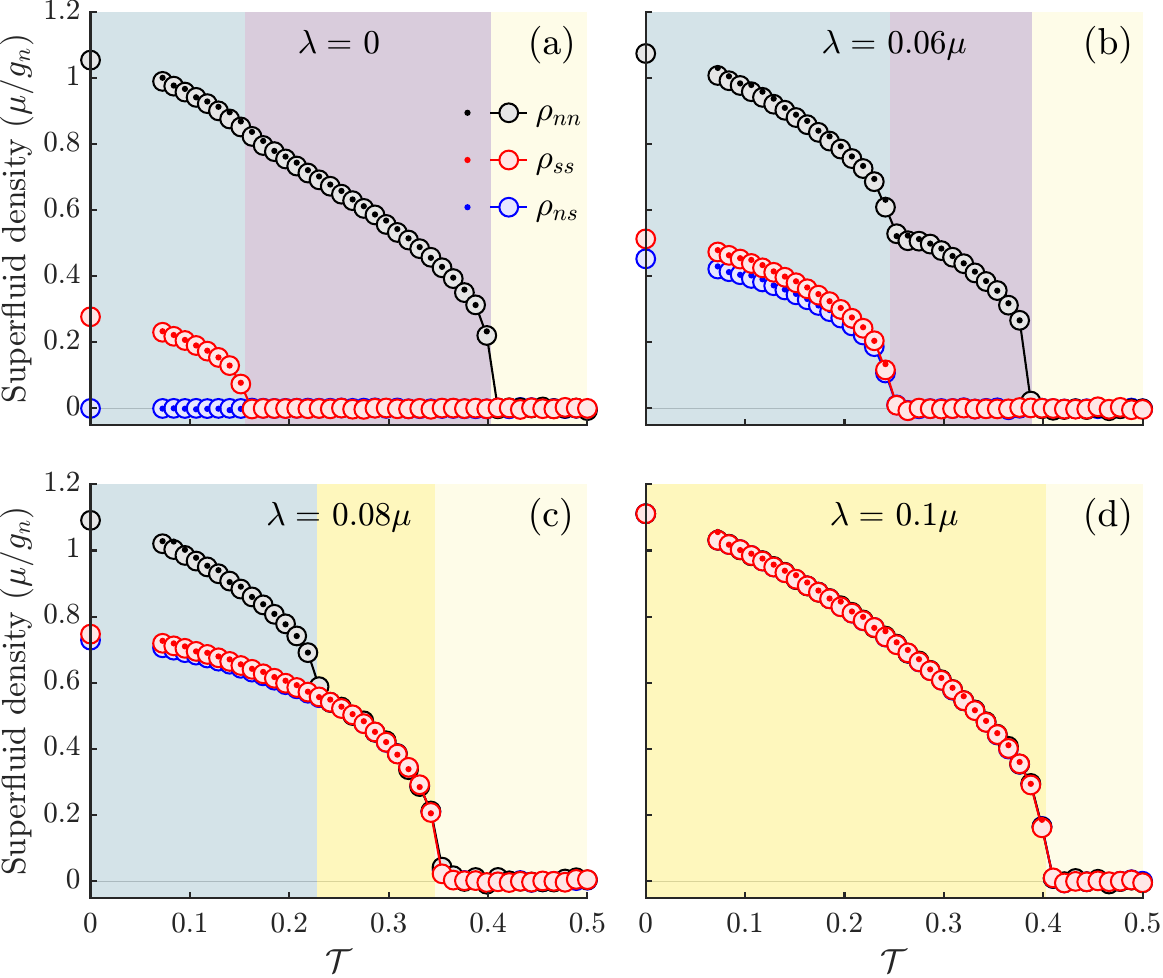}
	\caption{Temperature dependence of superfluid densities $\rho_{nn}$ (black), $\rho_{ss}$ (red) and $\rho_{ns}$ (blue) at (a) $\lambda=0$, (b) $\lambda=0.06\mu$, (c) $\lambda=0.08\mu$, and (d) $\lambda=0.1\mu$. Circles are evaluated via the procedure described in Appendix \ref{computesuperfluid}. Dots are evaluated via decay exponents $\eta_{\nu}$ extracted from fits to algebraically decaying correlations $G_{\nu}$ (see text). Zero temperature points are computed via Eq.~\eqref{eq:rhozeroT}. Background colors indicate superfluid phases from Fig.~\ref{fig:superfluiddensities}(d). All results are computed with $\mathcal{N}=512$.\label{fig:massspinsf}}
\end{figure}

Conversely, we may express the superfluid densities in terms of the decay exponents. In the easy-axis and mass superfluid phases this is achieved directly via Eqs.~(\ref{eq:eaeta}) and (\ref{eq:meta}). In the mass and spin superfluid phase, inverting Eq.~(\ref{eq:conciseeta}) gives
\begin{equation}
	\begin{split}
	\rho_{nn} &= \frac{Mk_{B}T}{2\pi\hbar^{2}\eta_0}\frac{1}{1-\mathcal{E}^2},\\
	\rho_{ss} &= \frac{Mk_{B}T}{2\pi\hbar^{2}\eta_\perp}\frac{1}{1-\mathcal{E}^2},\\
	\rho_{ns}&= \frac{Mk_{B}T}{2\pi\hbar^{2}\sqrt{\eta_\perp\eta_0}}\frac{\mathcal{E}}{1-\mathcal{E}^2},\label{eq:rhofrmeta_ns}
	\end{split}
\end{equation}
with $\mathcal{E}=\bar{\eta}/\sqrt{\eta_\perp\eta_0}$. In Fig.~\ref{fig:massspinsf} we compare superfluid densities extracted from equilibrium current-current correlations (circles) to those computed from Eqs.~(\ref{eq:eaeta}\,-\,\ref{eq:rhofrmeta_ns}), with decay exponents extracted from fits to correlations (\ref{eq:corr}) (dots). We find agreement at all temperatures for multiple values of $\lambda$. The relations (\ref{eq:rhofrmeta_ns}) illustrate the effect of $\rho_{ns}$ on the mass and spin superfluid densities. In particular, the emergence of $\rho_{ns}$ results in a two-step transition in the mass superfluidity, see Fig.~\ref{fig:massspinsf}(b) and (c).


\section{Conclusion}

In this paper we investigated the effect of axial magnetization on the superfluid properties of a ferromagnetic spin-1 Bose gas in two dimensions. This system supports superfluidity of both mass and spin currents, arising via respective BKT transitions. We find that the spin superfluid transition temperature increases with increasing magnetic potential, ultimately coinciding with the mass superfluid transition temperature. We thus identify three superfluid regimes: One with only mass superfluidity, one with distinct mass and spin superfluidity, and one with identical mass and spin superfluidity. We have quantified the interdependence of mass and spin currents at nonzero magnetic potential via the introduction of a third superfluid quantity, $\rho_{ns}$.

This quantity is analogous to the superfluid drag present in binary superfluids which, as with the system considered here, exhibit a $\mathrm{U}(1)\times\mathrm{U}(1)$ symmetry. However, we note an important distinction: The two superflows exhibited by the binary fluid correspond directly to transport of the individual fluid components, with superfluid drag describing entrainment between them. Contrarily, the two superflows exhibited by the broken-axisymmetric spin-1 Bose gas are associated with transport of mass and axial magnetization, with $\rho_{ns}$ quantifying the extent to which these superflows are identical; the mass and spin superfluid currents do not correspond to flow of distinct `mass' and `spin' fluid components.

Furthermore, we have presented the relationship between the relevant correlations and superfluidity in the presence of axial magnetization. Recent work on spatially resolved measurement of collective spin observables and the associated correlation functions in quasi-1D spin-1 gases~\cite{prufer_observation_2018,prufer_condensation_2022} should be applicable to the 2D regime considered here. Axial magnetization will also affect the nature of vortices in this system~\cite{lovegrove2012,lovegrove2016,PhysRevResearch.3.013154}; exploring this in the context of the BKT transitions identified in this work would be an interesting area to explore.

\section{Acknowledgments}
X.Y. acknowledges support from the National Natural Science Foundation of China (Grant No. 12175215), the National Key Research and Development Program of China (Grant No. 2022YFA 1405300) and  NSAF (Grant No. U2330401). A.P.C.U and P.B.B. acknowledge support from the Marsden Fund of the Royal Society of New Zealand. This research was supported by the Australian Research Council Centre of Excellence for Engineered Quantum Systems (EQUS, CE170100009). This research was partially supported by the Australian Research Council Centre of Excellence in Future Low-Energy Electronics Technologies (project number CE170100039) and funded by the Australian Government. The authors wish to acknowledge the use of New Zealand eScience Infrastructure (NeSI) high performance computing facilities as part of this research.


%

\appendix

\section{Computing superfluid densities}\label{computesuperfluid}

To compute the definitions (\ref{eq:sfdefns}) we consider the free energy $F=-k_{B}T\ln Z$, where the partition function
\begin{equation}
Z=\int\dd{\Psi}\exp\left[-\left(E-\mu N-\lambda M_{z}\right)/k_{B}T\right]\label{eq:partition}
\end{equation}
is evaluated over states $\Psi$ satisfying periodic boundary conditions, with the effect of the phase twist (\ref{eq:transform}) incorporated into the state energy via the transformation
\begin{align}
	E = E_0+\sum_{i,j}\frac{\hbar^{2}\kappa_{i}\kappa_{j}}{2M}N_{ij}+\hbar\int\dd{\mathbf{r}}\left(\kappa_{n}\mathbf{J}_{n}+\kappa_{s}\mathbf{J}_{s}\right)\cdot\hat{\mathbf{n}}\label{eq:Etransform}.
\end{align}
Here $E_0$ is the energy prior to the transformation (\ref{eq:energy}), $N_{ij}=\int\dd{\mathbf{r}}\Psi^{\dagger}A_{i}A_{j}\Psi$, $A_{n}=\mathbb{1}$, $A_{s}=f_{z}$ and $\mathbf{J}_i$ are the total currents (\ref{eq:totalcurrents}). Substituting Eq.~\eqref{eq:Etransform} into Eq.~\eqref{eq:partition} and computing Eq.~\eqref{eq:sfdefns} from the free energy gives
\begin{equation}
	\rho_{ij} = n_{ij} - \frac{M}{k_{B}TL^{2}}\int\dd{\mathbf{r}}\int\dd{\mathbf{r}'}\hat{\mathbf{n}}\cdot\expval{\mathbf{J}_{i}(\mathbf{r})\mathbf{J}_{j}(\mathbf{r}')}\cdot\hat{\mathbf{n}}\label{eq:rhoij}
\end{equation}
where $n_{ij}=\expval{N_{ij}}/L^{2}$, and $\mathbf{J}_{i}(\mathbf{r})\mathbf{J}_{j}(\mathbf{r}')$ denotes the outer product of $\mathbf{J}_{i}(\mathbf{r})$ and $\mathbf{J}_{j}(\mathbf{r}')$. Note that the mass and spin momenta are $P_{i} = M\int\dd{\mathbf{r}}\mathbf{J}_{i}\cdot\hat{\mathbf{n}}$ and hence Eq.~\eqref{eq:rhoij} can also be written as~\cite{PhysRevB.36.8343}
\begin{equation}
	\rho_{ij} = n_{ij} - \frac{\expval{P_{i}P_{j}}}{ML^{2}k_{B}T}.\label{eq:p2}
\end{equation}
In the infinite system size limit the current-current correlation function $\expval{\mathbf{J}_{i}(\mathbf{r})\mathbf{J}_{j}(\mathbf{r}')}$ depends only on the separation $\mathbf{r}-\mathbf{r}'$. With this one may express Eq.~(\ref{eq:rhoij}) as
\begin{equation}
	\rho_{ij} = n_{ij} -\lim_{L^{2}\to\infty}\int\dd{\mathbf{k}}\Delta(\mathbf{k})\left[\hat{\mathbf{n}}\cdot\chi_{ij}(\mathbf{k})\cdot\hat{\mathbf{n}}\right]\label{eq:rhoijcc}
\end{equation}
where $\Delta(\mathbf{k})\equiv \int\dd{\mathbf{r}}\frac{1}{4\pi^{2}}\mathrm{e}^{\mathrm{i}\mathbf{k}\cdot\mathbf{r}}$ and
\begin{equation}
	\chi_{ij}(\mathbf{k}) \equiv \frac{M}{k_{B}TL^{2}}\expval{\tilde{\mathbf{J}}_{i}(\mathbf{k})\tilde{\mathbf{J}}_{j}^{*}(\mathbf{k})},
\end{equation}
with $\tilde{\mathbf{J}}_{i}(\mathbf{k})=\int\dd{\mathbf{r}}\mathrm{e}^{-\mathrm{i}\mathbf{k}\cdot\mathbf{r}}\mathbf{J}_{i}(\mathbf{r})$.

The function $\Delta(\mathbf{k})$ approaches a delta-function in the infinite system size limit $L^{2}\to\infty$. However, the way in which this limit is taken is of critical importance. Consider the fluid to be confined within a rectangular box of dimensions $L_{n}\times L_{m}$ with walls oriented along perpendicular directions $\hat{\mathbf{n}}$ and $\hat{\mathbf{m}}$. To probe superfluidity, one must take the limit $L_{n}\to\infty$ first (see, for example \cite{PhysRevA.81.023623}). The function $\Delta(\mathbf{k})$ then acts to ensure the integration is performed over $\mathbf{k}\perp\hat{\mathbf{n}}$, so that $\hat{\mathbf{n}}\cdot\chi_{ij}(\mathbf{k})\cdot\hat{\mathbf{n}} = \chi_{ij}^{T}(|\mathbf{k}|)$ gives the transverse response. Alternatively, taking first $L_{m}\to\infty$ results in integration over $\mathbf{k}\parallel\hat{\mathbf{n}}$ so that $\hat{\mathbf{n}}\cdot\chi_{ij}(\mathbf{k})\cdot\hat{\mathbf{n}} = \chi_{ij}^{L}(|\mathbf{k}|)$ gives the longitudinal response, satisfying $\chi_{ij}^{L}(0)=n_{ij}$. With these considerations one obtains
\begin{equation}
	\rho_{ij} = \lim_{|\mathbf{k}|\to 0}\left[\chi_{ij}^{L}(|\mathbf{k}|)-\chi_{ij}^{T}(|\mathbf{k}|)\right].\label{eq:ltdiff}
\end{equation}

\begin{figure}
	\includegraphics[width=\linewidth]{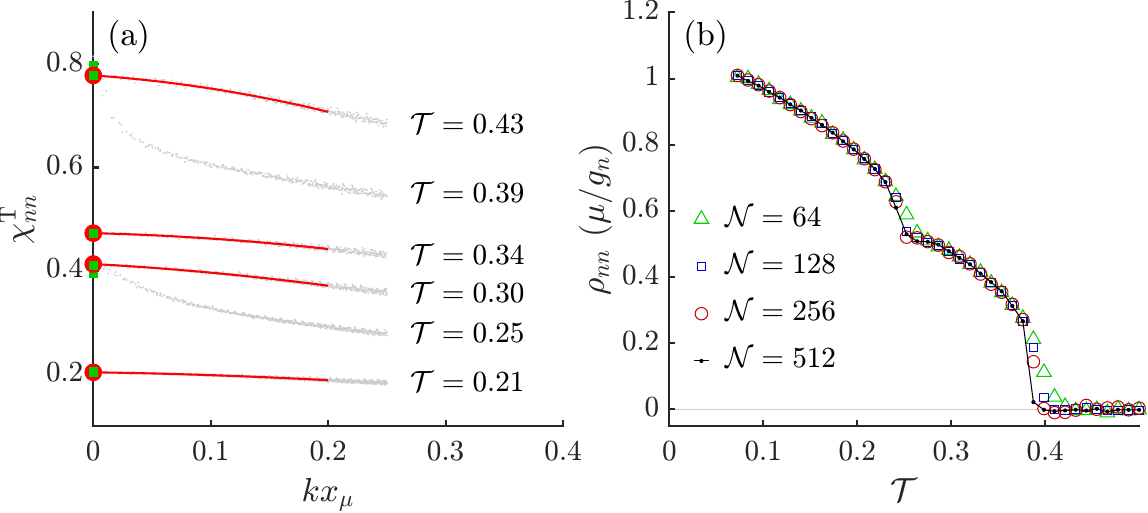}
	\caption{Attainment of mass superfluid density at $\lambda=0.06\mu$. (a) Wave-number dependence of transverse response function $\chi_{nn}^{\mathrm{T}}$ (gray dots). Red lines are fits used to extract the $k\to 0$ limit. Green squares indicate values computed via squared momentum expectations, see text. (b) Temperature dependence of mass superfluid density obtained with system sizes $\mathcal{N}=64$, $128$, $256$, and $512$.\label{fig:simdetails}}
\end{figure}

We compute superfluid densities via Eq.~(\ref{eq:ltdiff}). Away from transition temperatures a quadratic fit to the response functions $\chi^{L}_{ij}(|\mathbf{k}|)$ and $\chi^{T}_{ij}(|\mathbf{k}|)$ is sufficient to extract the $|\mathbf{k}|\to 0$ limit, see Fig.~\ref{fig:simdetails}(a). However, near transition temperatures the response function $\chi^{T}_{ij}(|\mathbf{k}|)$ changes rapidly near $\mathbf{k}=\mathbf{0}$, and the fitting procedure becomes unreliable. We therefore compute the $|\mathbf{k}|\to 0$ limit via $\lim_{|\mathbf{k}|\to 0}\chi^{T}_{ij}(|\mathbf{k}|)\to \langle P_{i}P_{j}\rangle/ML^{2}k_{B}T$, c.~f. Eq.~(\ref{eq:p2}).

We have computed superfluid densities on $\mathcal{N}\times \mathcal{N}$ point numerical grids with $\mathcal{N}=64$, $128$, $256$, and $512$. Example results are shown in Fig.~\ref{fig:simdetails}(b). System-size dependence is observed near superfluid transition temperatures.

\end{document}